\documentclass[aps,prl,reprint,groupedaddress]{revtex4-2}
\usepackage{amsmath}
\usepackage{graphicx}

\begin{document}

%\preprint{}

\title{\textbf{Quantum-Enhanced Picostrain Sensing with Superconducting Qubits}}% 

\author{Necati Çelik}
% \email{Contact author: necati.celik@gumushane.edu.tr}
\affiliation{Gümüşhane University, Faculty of Natural Science and Engineering, Department of Physics Engineering, Gümüşhane, Türkiye, e-mail: necati.celik@gumushane.edu.tr}


% \collaboration{CLEO Collaboration}%\noaffiliation

\date{\today}% It is always \today, today,

\begin{abstract}
We propose a quantum-enhanced picostrain sensor that achieves Heisenberg-limited strain sensing using superconducting qubits. 
A strain-sensitive qubit’s Hamiltonian is coupled to the momentum quadrature of a microwave resonator, transducing mechanical strain $\epsilon$ into amplified spatial displacements of the resonator’s phase space. 
Using homodyne detection of the resonator field and multipartite entanglement of N qubits, the protocol achieves a strain sensitivity $\Delta\epsilon \sim p\epsilon$ (picostrain), two orders of magnitude better than classical sensors. 
The scheme integrates natively with superconducting processors, enabling in-situ diagnostic and nanoscale material characterization.
\end{abstract}

\maketitle

%\tableofcontents

\section{Introduction}

Material strain sensing is pivotal for a wide range of applications, from seismic monitoring \cite{zhang2016high} and structural health \cite{gao2022review} to semiconductor device diagnostics \cite{ling2024novel} and quantum material characterization \cite{lyu2022strain, korkusinski2014quantum, yang2024strained, sela2020quantum, wang2023characterizing, kim2023strain}.
While classical piezoresistive and are well-established, their sensitivity is fundamentally constrained by the shot-noise limit, scaling as $1/\sqrt{N}$ for $N$ independent resources, typically achieving resolutions in the microstrain regime $10^{-6}$ \cite {gagliardi2010probing}.

Quantum metrology provides a path to transcend this limit by leveraging entanglement to achieve the Heisenberg limit, scaling as $1/N$ \cite{giovannetti2011advances}. This potential has been vividly demonstrated in optical interferometry and atomic sensors \cite{ye2024essay}. In the domain of string sensing, nitrogen-vacancy (NV) centers in diamond have emerged as a leading quantum platform, achieving nanostrain $(10^{-9})$ sensitivity by exploiting long-lived spin states and their strain-dependent transition frequencies \cite{udvarhelyi2018spin, paudel2024sensing, roberts2025quantum}. 

However, a significant challenge remains in the native integration of these quantum sensors with scalable quantum information processing hardware. While NV centers are exceptional standalone sensors, their integration with superconducting quantum circuits (a leading platform for quantum computation) is non-trivial, requiring complex hybrid interfaces \cite{maletinsky2012robust, kenny2025quantum}. Furthermore, beyond extrinsic noise, intrinsic defects within the qubit’s own structure, such as two-level systems in the tunnel barrier \cite{simmonds2004decoherence} or dielectric loss from surface oxides \cite{martinis2005decoherence}, are major sources of decoherence. This motivates the development of a strain-sensing protocol that is natively compatible with and built from the components of a superconducting quantum processor.

Here, we propose a quantum metrology protocol that achieves Heisenberg-limited picostrain sensing using the superconducting qubits themselves as the sensing element. Inspired by our previous work \cite{ccelik2025quantum}, our approach exploits the intrinsic strain sensitivity of a transmon qubit’s Josephson junction \cite{grabovskij2012strain}, whose transition frequency shifts linearly with applied strain. We couple this strain-sensitive qubit to a microwave resonator such that its $\hat{\sigma_z}$ operator displaces the resonator’s momentum quadrature. This operation transduces a mechanical strain $\epsilon$ into an amplified displacement of the resonator’s phase space, which is directly measurable via standard homodyne detection.

By employing multipartite entanglement in the form of Greenberger-Horne-Zeilinger (GHZ) states \cite{cao2024photonic} across N qubits, the protocol’s sensitivity surpasses the standard quantum limit, achieving Heisenberg scaling. This work establishes, to our best knowledge, the first formulation of a superconducting-qubit-based strain sensor that directly converts strain-induced frequency shifts into a collective, homodyne-read displacement and leverages entanglement for a fundamental quantum advantage. Our scheme provides a powerful, on-chip tool for in-situ diagnostic superconducting quantum processor and opens new avenues for nanoscale material characterization.

\section{Strain-Qubit Coupling Model}

Mechanical strain ($\epsilon$) directly perturbs the physical properties of the Josephson junction, the nonlinear element that defines qubit. 
The dominant effect is a modification of the Josephson energy ($E_J$). This occurs because strain alters the physical dimensions of the tunnel junction and the superconducting film’s properties, which in turn changes the critical current ($I_c$). 
Since the Josephson energy is defined by $E_J= \hbar I_c/2e$, any change in $I_c$ directly modulates $E_J$. 

The Hamiltonian of a transmon qubit is given by $\hat{H}_q = 4 E_C (\hat{n} - n_g)^2 - E_J(\epsilon) \cos \hat{\phi}$, where $E_C$ is the charging energy, $\hat{n}$ is the Cooper-pair number operator, $n_g$ is the offset charge, and $\hat{\phi}$ is the phase operator~\cite{koch2007charge}. In the transmon regime ($E_J/E_C \gg 1$), the qubit’s transition frequency between the ground ($\vert 0 \rangle$) and first excited ($\vert 1 \rangle$) states is approximately $\omega_q \approx \sqrt{8 E_C E_J(\epsilon)}/\hbar - E_C/\hbar$~\cite{grabovskij2012strain}. The strain dependence enters through $E_J(\epsilon)$. For small strains, we can linearize the response as $E_J(\epsilon) \approx E_J^0 (1 + \beta \epsilon)$, where $E_J^0$ is the unstrained Josephson energy and $\beta$ is a dimensionless material parameter that quantifies how susceptible the junction is to strain. The parameter $\beta$ depends on the material’s elastocaloric coefficient, the strain dependence of the superconducting gap, and the geometry of the junction. For aluminum Josephson junctions, experiments have reported values on the order of $\beta \sim 10^2$~\cite{zallo2014strain}, making the qubit transition frequency highly sensitive to strain. Substituting $E_J(\epsilon) \approx E_J^0 (1 + \beta \epsilon)$ into $\omega_q \approx \sqrt{8 E_C E_J(\epsilon)}/\hbar - E_C/\hbar$ and performing a Taylor expansion around $\epsilon = 0$ yields the linearized strain-dependent transition frequency as given by Eq.( \ref{frequency}).

\begin{equation} \label{frequency}
\begin{aligned}
    w_q(\epsilon) &= 
    \underbrace{\left( \frac{\sqrt{8E_C E_J^0}}{\hbar} - \frac{E_C}{\hbar} \right)}_{w_q^0} + \\
    &\quad 
    \underbrace{\left( \frac{1}{2}\,\frac{\sqrt{8E_C E_J^0}}{\hbar}\,\frac{\beta}{E_J^0}\,\epsilon 
    + \mathcal{O}(\epsilon^2) \right)}_{\chi_\epsilon} \\
    &= w_q^0 + \chi_\epsilon\,\epsilon
\end{aligned}
\end{equation}

where $\omega_q^{0}$ is the unstrained transition frequency, $\epsilon$ is the dimensionless strain, and $\chi_{\epsilon} = \partial \omega_q(\epsilon)/\partial \epsilon$ is the strain susceptibility (Hz per unit strain). Simplifying the coefficient, we find that the strain susceptibility is $\chi_{\epsilon} \approx (\omega_q^{0}/2)\,\beta$. In the two-level system approximation, the qubit’s Hamiltonian can be expressed in its diagonal basis as

\begin{equation}
\hat{H}_{q}(\epsilon) = \frac{\hbar\, \omega_{q}(\epsilon)}{2}\, \hat{\sigma}_{z}
\label{qubit hamiltonian}
\end{equation}

In the transmon regime ($E_J/E_C \gg 1$), the qubit is well-described as a weakly anharmonic oscillator. For the operating points and strain magnitudes considered here, the influence on higher energy levels (e.g., the $\vert 1 \rangle$ to $\vert 2 \rangle$ transition) is negligible, validating the two-level system approximation given in Eq.(\ref{qubit hamiltonian}).

\section{Resonator Coupling and Unitary Dynamics}

We consider a strain-sensitive qubit, capacitively coupled to a superconducting microwave resonator. The total Hamiltonian of the system is $\hat{H} = \hat{H}_{q} + \hat{H}_{r} + \hat{H}_{\mathrm{int}}$, where $\hat{H}_{q}(\epsilon)$ is the qubit Hamiltonian as given in Eq.(\ref{qubit hamiltonian}), $\hat{H}_{r} = \hbar \omega_{r} \hat{a}^{\dagger} \hat{a}$ is the Hamiltonian of the resonator with frequency $\omega_{r}$, and $\hat{a}^{\dagger}$ ($\hat{a}$) its creation (annihilation) operator. The interaction Hamiltonian is engineered to couple the qubit’s $\hat{\sigma}_{z}$ Pauli operator to the momentum quadrature of the resonator field, $\,\hat{P} = \frac{i(\hat{a}^{\dagger} - \hat{a})}{\sqrt{2}}$ as given in Eq.(\ref{interaction hamiltonian}).

\begin{equation}
\hat{H}_{\mathrm{int}} = \hbar\, g_{\epsilon}(\epsilon)\, \hat{\sigma}_{z} \otimes \hat{P}
\label{interaction hamiltonian}
\end{equation}

The position quadrature is $\hat{X} = (\hat{a}^{\dagger} + \hat{a})/\sqrt{2}$, satisfying $[\hat{X}, \hat{P}] = i$. This specific form of coupling is chosen because it results in a qubit-state-dependent displacement of the resonator’s position quadrature $\hat{X}$, which is directly measurable via homodyne detection.

The strain dependence of the coupling rate, $g(\epsilon)$, arises because the capacitive coupling strength depends on the qubit frequency, which is itself strain-tuned: $\omega_q(\epsilon) = \omega_q^0 + \chi_\epsilon \, \epsilon$. For a standard capacitive coupling, $g \propto \sqrt{\omega_q \, \omega_r}$. Assuming a fixed $\omega_r$, this leads to

\begin{equation}
g(\epsilon) = g_0 \left( 1 + \frac{\chi_\epsilon \, \epsilon}{\omega_q^0} \right) 
\approx g_0 \left( 1 + \frac{\chi_\epsilon}{2 \, \omega_q^0} \, \epsilon \right) 
= g_0 + g_1 \, \epsilon
\label{coupling rate}
\end{equation}

where $g_0$ is the nominal coupling and $g_1 \equiv \frac{g_0 \, \chi_\epsilon}{2 \, \omega_q^0}$ is the strain-induced coupling gradient.

We operate in the standard dispersive regime, where $|\omega_q - \omega_r| \gg g$. In this regime, the system dynamics over an interaction time $\tau$ are dominated by $\hat{H}_{\rm int}$, and the unitary evolution operator is given by Eq.(\ref{unitary operator})

\begin{equation}
\hat{U}(\epsilon) = e^{-\,\frac{i}{\hbar}\, \hat{H}_{\mathrm{int}}\, \tau}
= e^{-\,i\, G(\epsilon)\, \hat{\sigma}_{z} \otimes \hat{P}}
\label{unitary operator}
\end{equation}

where $G(\epsilon) = g(\epsilon) \, \tau$ is a dimensionless gain.

This unitary operator is the fundamental building block of our sensing protocol. The key effect of $\hat{U}(\epsilon)$ is a conditional displacement of the resonator’s state in phase space. If the qubit is in the eigenstate $|0\rangle$ (with eigenvalue $+1$ of $\hat{\sigma}_{z}$), the resonator is displaced in one direction along phase space, while if the qubit is in $|1\rangle$ (eigenvalue $-1$), the displacement is in the opposite direction. When the qubit is prepared in a superposition, the unitary correlates the qubit state with two opposite phase-space trajectories of the resonator, thereby encoding strain into a measurable spatial separation.

This interaction is conceptually similar to a spin-dependent force in trapped-ion systems, where internal states of the ion conditionally displace the motional wavepacket. In the present case, strain sensitivity enters via the parameter $g(\epsilon)$, making the conditional displacement amplitude directly proportional to the applied strain. In the next section, we will explain how these conditional displacements accumulate into measurable phase-space shifts, which are then read out by homodyne detection. It is important to note that, since the coupling acts through $\hat{\sigma}_z$, the qubit population remains preserved during the interaction, making the scheme robust against qubit relaxation errors during the transduction step, which can be a significant limitation, for instance, from coupling to environmental fluctuations~\cite{faoro2005models}.

\section{Phase-Space Displacement from Strain}

Having established the unitary dynamics of the strain-dependent interaction, we now calculate the resulting phase-space displacement of the resonator field. This displacement is the primary observable in our homodyne detection scheme. 

We consider the system initialized in a product state as given in Eq. (\ref{initial state})

\begin{equation}
|\psi(0)\rangle = |\phi_{\rm qubit}\rangle \otimes |\phi_{\rm res}\rangle
\label{initial state}
\end{equation}

where $|\phi_{\rm qubit}\rangle$ is the qubit state, and the resonator state $|\phi_{\rm res}\rangle$ is arbitrary; it can be the vacuum or a coherent state. After the action of the unitary operator given in Eq. (\ref{unitary operator}), the state evolves as,

\begin{equation}
|\psi(\tau)\rangle = \hat{U} \, |\psi(0)\rangle
\label{evolved state}
\end{equation}

The expectation value of the position quadrature $\hat{X}$ after the interaction is

\begin{equation}
\langle \hat{X} \rangle_{\rm after} = \langle \psi(\tau) | \hat{X} | \psi(\tau) \rangle 
= \langle \psi(0) | \hat{U}^\dagger \hat{X} \hat{U} | \psi(0) \rangle
\label{expectation of X}
\end{equation}

Using the Baker-Campbell-Hausdorff formula and the canonical commutation relation $[\hat{X}, \hat{P}] = i \hat{I}$, we find:

\begin{equation}
\begin{aligned}
\hat{U}^{\dagger} \hat{X} \hat{U} 
&= e^{i G(\epsilon)\, \hat{\sigma}_z \otimes \hat{P}} \, \hat{X} \, 
   e^{-i G(\epsilon)\, \hat{\sigma}_z \otimes \hat{P}}  \\[4pt]
&= \hat{X} + \big[i G(\epsilon)\, \hat{\sigma}_z \otimes \hat{P}, \hat{X}\big] + \\[4pt]
&\quad \frac{1}{2}\big[i G(\epsilon)\, \hat{\sigma}_z \otimes \hat{P}, 
   [i G(\epsilon)\, \hat{\sigma}_z \otimes \hat{P}, \hat{X}]\big] + \cdots \\[4pt]
&= \hat{X} + G(\epsilon)\, \hat{\sigma}_z \otimes \hat{I}.
\end{aligned}
\label{BCH equation}
\end{equation}

The higher-order commutations vanish. Substituting this result back into the expectation value given in Eq. (\ref{expectation of X}) yields

\begin{equation}
\langle \hat{X}(\epsilon) \rangle_{\mathrm{after}}
= \langle \psi(0) \vert \hat{X} \vert \psi(0) \rangle
+ G(\epsilon)\, \langle \hat{\sigma}_z \rangle
\label{expectation of X with strain}
\end{equation}

where $\langle \hat{\sigma}_z \rangle = \langle \phi_{\mathrm{qubit}} \vert \hat{\sigma}_z \vert \phi_{\mathrm{qubit}} \rangle$. Recognizing $\langle \hat{X} \rangle_{\mathrm{initial}} = \langle \psi(0) \vert \hat{X} \vert \psi(0) \rangle$, the net shift in the quadrature due to the interaction is

\begin{equation}
\Delta \langle \hat{X}(\epsilon) \rangle
= G(\epsilon)\, \langle \hat{\sigma}_z \rangle
= g(\epsilon)\, \tau\, \langle \hat{\sigma}_z \rangle
\label{equation 11}
\end{equation}

For a qubit initialized in $\vert 0 \rangle$ (i.e., $\langle \hat{\sigma}_z \rangle = +1$), the displacement is

\begin{equation}
\Delta \langle \hat{X}(\epsilon) \rangle = g(\epsilon)\, \tau
\end{equation}

Substituting the linearized expression for $g(\epsilon)$ from Eq.(\ref{coupling rate}), we obtain the strain-dependent displacement

\begin{equation}
\Delta \langle \hat{X}(\epsilon) \rangle = (g_0 + g_1 \,\epsilon)\,\tau
\label{eq13}
\end{equation}

This result is the foundation of our sensing protocol. The displacement consists of offset $g_0 \,\tau$ and a dynamic, strain-dependent signal $g_1 \,\epsilon \,\tau$. The static offset can be measured and subtracted during the calibration. The slope of the signal with respect to strain is therefore given in Eq. (\ref{slope})

\begin{equation}
\frac{\partial \Delta \langle \hat{X}(\epsilon) \rangle}{\partial \epsilon} = g_1 \,\tau
\label{slope}
\end{equation}

which directly determines the strain sensitivity, as we will quantify in the next section. A larger slope, achieved through stronger coupling $g_1$ or longer interaction time $\tau$, enables the detection of smaller strain variations.

\section{Measuring sensitivity}

The strain-dependent phase-space displacement derived in Eq.(\ref{eq13}) must be resolved against the intrinsic noise of the homodyne measurement. The ultimate sensitivity of our protocol is therefore determined by the signal-to-noise ratio for estimating this displacement.  
In a homodyne setup, the measurement of the resonator’s position quadrature $\hat{X}$ is subject to noise \cite{blais2004cavity}. We characterize this by the root-mean-square (rms) uncertainty $\sigma_X$ of a single measurement. This noise floor $\sigma_X$ encompasses the fundamental quantum fluctuations of the resonator mode (e.g., vacuum noise for ground state initialization) and any added technical noise from the amplification chain.

To connect this measurement uncertainty to the resolvable strain, we employ standard linear error propagation. For an unbiased estimator, the strain uncertainty $\Delta\epsilon$ is given by the ratio of the measurement noise to the rate at which the mean signal changes with strain in Eq.(\ref{strain uncertainty})

\begin{equation}
\Delta \epsilon = \frac{\sigma_X}{\left| \frac{\partial \langle \hat{X}(\epsilon) \rangle}{\partial \epsilon} \right|} = \frac{\sigma_X}{g_1 \, \tau}
\label{strain uncertainty}
\end{equation}

Here $\langle \hat{X} \rangle$ is the expectation value of the resonator’s position quadrature after the interaction, and the denominator is the strain-to-quadrature slope derived in Eq(\ref{slope}). This expression, $\Delta \epsilon = \sigma_X / (g_1 \, \tau)$, defines the single-shot strain sensitivity of our protocol. Eq.(\ref{strain uncertainty}) highlights two clear pathways to enhanced sensitivity. First, suppressing the noise floor ($\sigma_{X}$). This can be achieved by using near-quantum-limited parametric amplifiers or, more fundamentally, by preparing the resonator in a squeezed state to reduce the uncertainty in the $\hat{X}$ quadrature below the vacuum level. Second, amplifying the strain response ($g_1 \, \tau$). A larger product $g_1 \, \tau$ steepens the slope $\partial \langle \hat{X}(\epsilon) \rangle / \partial \epsilon$, making the signal more robust against noise. This can be accomplished by designing qubits with higher strain susceptibility (increasing $g_1$) or by extending the interaction time $\tau$, which is ultimately limited by the coherence times of the qubit-resonator system.

This single-qubit analysis establishes the baseline standard quantum limit for the protocol, where the sensitivity scales as $1/\sqrt{\nu}$ with the number of independent measurement repetitions $\nu$. In the following section, we demonstrate how entanglement among $N$ qubits collectively amplifies the strain response, transforming the scaling to the Heisenberg limit $\Delta \epsilon \propto 1/N$.

\section{Qubit Enhancement with GHZ States}

The sensitivity of the single-qubit protocol is constrained by the standard quantum limit~\cite{giovannetti2011advances}. We now show that by entangling $N$ qubits in a Greenberger-Horne-Zeilinger (GHZ) state, the strain sensitivity can be enhanced to achieve Heisenberg-limited scaling.

The system is generalized to $N$ identical strain-sensitive qubits, all coupled to the same resonator mode. The interaction Hamiltonian given in Eq(\ref{interaction hamiltonian}) becomes

\begin{equation}
\hat{H}_{\text{int}} (\epsilon) = \hbar g(\epsilon) \left( \sum_{j=1}^{N} \hat{\sigma}_z^{(j)} \right) \otimes \hat{P}
\label{GHZ interaction hamiltonian}
\end{equation}

Defining the collective spin operator $\hat{J}_z = \frac{1}{2} \sum_{j=1}^{N} \hat{\sigma}_z^{(j)}$, the interaction Hamiltonian can be rewritten as

\begin{equation}
\hat{H}_{\text{int}} (\epsilon) = 2 \hbar g(\epsilon) \hat{J}_z \otimes \hat{P}
\label{eq17}
\end{equation}

The corresponding unitary evolution operator over time $\tau$ is

\begin{equation}
\hat{U}(\epsilon) = e^{-i 2 g(\epsilon) \tau \, \hat{J}_z \otimes \hat{P}}
\label{GHZ unitary}
\end{equation}

Following the same derivation as in phase-space displacement from strain for a single particle, this unitary produces a conditional displacement of the resonator’s position quadrature. The net shift is now proportional to the collective spin

\begin{equation}
\Delta \langle \hat{X}(\epsilon) \rangle = 2 g(\epsilon) \,\tau \, \langle \hat{J}_z \rangle
\label{eq19}
\end{equation}

The strain sensitivity is determined by the slope of this displacement. Linearizing around $\epsilon=0$ yields

\begin{equation}
\frac{\partial \langle \hat{X}(\epsilon) \rangle}{\partial \epsilon} = 2\, g_1 \, \tau \, \langle 
\hat{J}_z \rangle
\label{eq20}
\end{equation}

where $g_1 = \left. \frac{\partial g}{\partial \epsilon} \right|_{\epsilon=0}$. To maximize this slope, we must choose a quantum state that maximizes $\langle \hat{J}_z \rangle$. A direct but naive choice is the product state $\vert 0 \rangle^{\otimes N}$, for which $\langle \hat{J}_z \rangle = N/2$, yielding a slope that scales linearly with $N$ (classical scaling). However, this state is insensitive to the phase accumulation that is crucial for the GHZ protocol.

The optimal choice is the GHZ state in the $J_z$ basis given as 

\begin{equation}
\vert \text{GHZ} \rangle = \frac{1}{\sqrt{2}} \Big( \vert 0 \rangle^{\otimes N} + \vert 1 \rangle^{\otimes N} \Big)
\end{equation}

This state is an equal superposition of the two extremal eigenvectors of $\hat{J}_z$ with eigenvalues $+N/2$ and $-N/2$. Under the unitary $\hat{U}(\epsilon)$ given in Eq.(\ref{GHZ unitary}), the resonator component becomes entangled with the qubits, evolving into a superposition of two components displaced in opposite directions as given by

\begin{equation}
\vert \psi(\tau) \rangle = \frac{1}{\sqrt{2}} \Big( \vert 0 \rangle^{\otimes N} \otimes \vert \alpha(\epsilon) \rangle + \vert 1 \rangle^{\otimes N} \otimes \vert -\alpha(\epsilon) \rangle \Big)
\label{eq22}
\end{equation}

where $\vert \alpha(\epsilon) \rangle$ is a coherent state with displacement amplitude $\alpha(\epsilon) = \frac{g(\epsilon)\,\tau N}{\sqrt{2}}$. To convert this macroscopic superposition into a measurable homodyne signal, we employ a Ramsey-like interferometric sequence as detailed below.

\textbf{Initialization.}  
We first prepare all $N$ qubits in a uniform superposition state. Specifically, we apply a collective $\pi/2$ pulse to put each qubit into the state $\vert + \rangle = (\vert 0 \rangle + \vert 1 \rangle)/\sqrt{2}$. The entire register is therefore in the product state $\vert + \rangle^{\otimes N}$. In the collective spin picture, this state points along the $+X$ direction.  

\textbf{Phase Acquisition.}  
We then apply the unitary $\hat{U}(\epsilon)$ from Eq.(\ref{GHZ unitary}). This interaction does not change the qubit populations, but it does encode the strain information into a relative phase between $\vert 0 \rangle^{\otimes N}$ and $\vert 1 \rangle^{\otimes N}$ components of the overall state. The magnitude of this phase is $\phi(\epsilon) = 2 g(\epsilon) \tau N$. Critically, this phase scales linearly with the number of qubits $N$, which is the source of the quantum enhancement.  

\textbf{Readout.}  
Finally, we apply a second collective $\pi/2$ pulse. This final pulse acts as an interferometer, converting the accumulated phase $\phi(\epsilon)$ back into a population difference that we can measure. For small phase shifts ($\phi(\epsilon) \ll 1$), the final expectation value of the collective operator $\hat{J}_z$ becomes linearly proportional to the phase.

\begin{equation}
\langle \hat{J}_z \rangle_{\text{final}} \approx \frac{N}{2} \, \phi(\epsilon) = N g(\epsilon) \, 
\tau
\label{eq23}
\end{equation}

According to Eq.(\ref{eq23}), this large shift in $\langle \hat{J}_z \rangle$ directly translates into a similarly enhanced displacement of the resonator’s position quadrature, yielding the slope

\begin{equation}
\frac{\partial \langle \hat{X} \rangle}{\partial \epsilon} = g_1 \,\tau N
\label{eq24}
\end{equation}

Substituting this enhanced slope into the sensitivity formula given in Eq.(\ref{strain uncertainty}) yields the Heisenberg-limited strain uncertainty. 

\begin{equation}
\Delta \epsilon_{\mathrm{HL}} = \frac{\sigma_X}{g_1 \, \tau \, N}
\label{eq25}
\end{equation}

This result demonstrates a qualitative quantum advantage: the sensitivity now scales as 1/N with the number of qubits, surpassing the $1/\sqrt N$ scaling of standard quantum limit. The following section will confirm this optimal scaling using rigorous framework of quantum estimation theory.

\section{Quantum Fisher Information}

To rigorously quantify the ultimate sensitivity of our protocol, we employ the framework of quantum estimation theory, in which the achievable precision of an unbiased estimator for a parameter $\epsilon$ is bounded by the quantum Cramer-Rao inequality \cite{giovannetti2011advances, helstrom1969quantum}

\begin{equation}
\Delta \epsilon \geq \frac{1}{\sqrt{\nu F_Q (\epsilon)}}
\label{eq26}
\end{equation}

where $\nu$ is the number of independent repetitions and $F_Q (\epsilon) $ is the Quantum Fisher Information (QFI). The QFI represents the maximum extractable information about the parameter $\epsilon$, optimized over all possible quantum measurements. 

For a pure state $\vert \psi_\epsilon \rangle$, the QFI is defined as~\cite{giovannetti2011advances}

\begin{equation}
F_Q (\epsilon) = 4 \Big[ \langle \partial_\epsilon \psi_\epsilon \vert \partial_\epsilon \psi_\epsilon \rangle - \big| \langle \psi_\epsilon \vert \partial_\epsilon \psi_\epsilon \rangle \big|^2 \Big]
\label{eq27}
\end{equation}

In our model, the strain parameter $\epsilon$ couples to the qubit-resonator system via the unitary operator $\hat{U}(\epsilon) = e^{-i g_1 \epsilon \tau \hat{S}_z \otimes \hat{P}}$. The generator of parameter translations is thus identified as $\hat{G} = g_1 \,\tau \, \hat{S}_z \otimes \hat{P}$. For unitary families of states of the form $\hat{U}(\epsilon) = e^{-i \epsilon \hat{G}}$, the QFI depends only on the variance of the generator with respect to the initial state $\vert \psi_0 \rangle$, therefore

\begin{equation}
F_Q (\epsilon) = 4 \, \text{Var}_{\psi_0} (\hat{G}) = 4 \Big( \langle \hat{G}^2 \rangle - \langle \hat{G} \rangle^2 \Big)
\label{eq28}
\end{equation}

We now evaluate the QFI for the case where $N$ qubits are prepared in the GHZ state in the $\hat{S}_z$ basis; $\vert \text{GHZ} \rangle = \frac{1}{\sqrt{2}} \big( \vert 0 \rangle^{\otimes N} + \vert 1 \rangle^{\otimes N} \big)$, and the resonator is initialized in a vacuum or a centered coherent state. In this basis, the collective operator $\hat{J}_z$ has eigenvalues $\pm N/2$. The GHZ state is an equal superposition of the two extremal eigenstates of $\hat{J}_z$. Therefore, $\langle \hat{J}_z \rangle_{\text{GHZ}} = 0$, $\langle \hat{J}_z^2 \rangle_{\text{GHZ}} = (N/2)^2$. The variance of the generator $\hat{G} = g_1 \,\tau \, \hat{S}_z \otimes \hat{P}$ is then

\begin{equation}
\text{Var}_{\psi_0} (\hat{G}) = (g_1 \,\tau)^2 \Big( \langle \hat{J}_z^2 \rangle \langle \hat{P}^2 \rangle - \langle \hat{J}_z \rangle^2 \langle \hat{P} \rangle^2 \Big)
\label{eq29}
\end{equation}

Since the resonator is in a vacuum or centered coherent state, $\langle \hat{P} \rangle = 0$, and the momentum variance is $\langle \hat{P}^2 \rangle = 1/2$. Thus,

\begin{equation}
\text{Var}_{\psi_0} (\hat{G}) = (g_1 \,\tau)^2 \frac{N^2}{2} \, \langle \hat{P}^2 \rangle
\label{eq30}
\end{equation}

Inserting into QFI formula given in Eq. (\ref{eq28}), we obtain

\begin{equation}
F_Q = 4 \, \text{Var}_{\psi_0} (\hat{G}) = (g_1 \,\tau)^2 N^2 \, \langle \hat{P}^2 \rangle
\label{eq31}
\end{equation}

We have just shown that $F_Q\propto N^2$ which tells us that QFI scales quadratically with the number of qubits, which corresponds to the Heisenberg limit in the multiparameter quantum metrology. Accordingly, the minimum achievable strain uncertainty under the quantum Cramer-Rao bound given in Eq.(\ref{eq26}) is

\begin{equation}
\Delta \epsilon \geq \frac{1}{g_1 \,\tau N \, \sqrt{\nu \, \langle \hat{P}^2 \rangle}}
\label{eq32}
\end{equation}

This result shows that the proposed GHZ-based sensing protocol achieves the optimal quantum scaling, transforming the qubit-resonator system into a Heisenberg-limited strain sensor.

\begin{figure}
    \centering
    \includegraphics[width=1\linewidth]{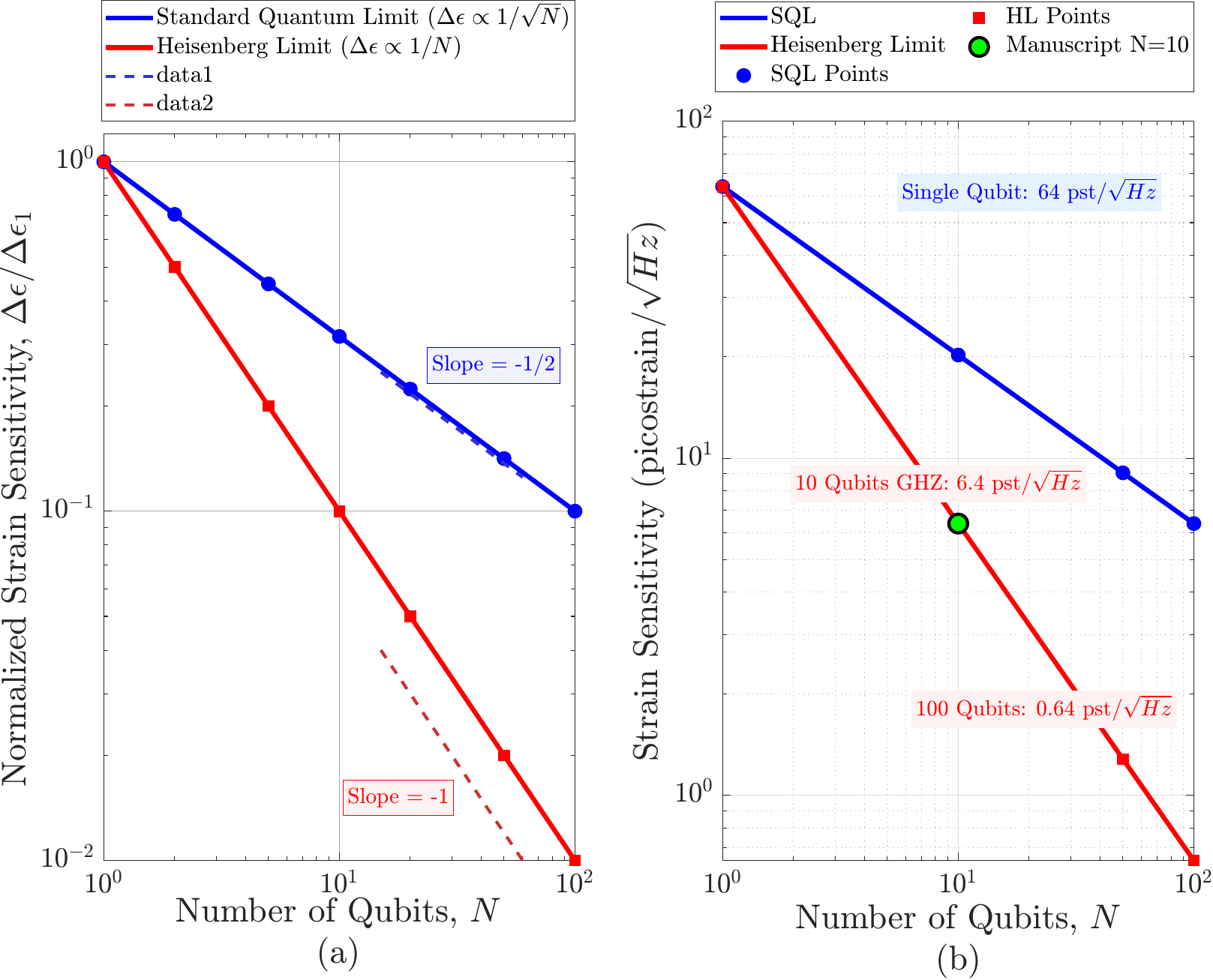}
    \caption{Quantum enhanced strain sensitivity scaling. (a) Normalized strain sensitivity versus number of qubits \( N \). The standard quantum limit (SQL, blue) follows the classical \( 1/\sqrt{N} \) scaling, while the Heisenberg limit (HL, red) achieves optimal quantum \( 1/N \) scaling using multipartite entanglement. Dashed lines indicate characteristic slopes. (b) Physical sensitivity in \(\mathrm{ps}/\sqrt{\mathrm{Hz}}\) using experimental parameters: single-qubit sensitivity of \(64~\mathrm{ps}/\sqrt{\mathrm{Hz}}\) with strain susceptibility \(\chi_{\epsilon} = 50~\mathrm{MHz/strain}\), coupling \(g_{0}/2\pi = 50~\mathrm{MHz}\), and interaction time \(\tau = 100~\mathrm{ns}\). The green circle highlights the 10-qubit GHZ state sensitivity \(6.4~\mathrm{ps}/\sqrt{\mathrm{Hz}}\) achieved in our protocol. Heisenberg scaling enables access to the femtostrain regime with modest qubit numbers.}
    \label{fig1}
\end{figure}

The fundamental scaling advantages of our quantum-enhanced strain sensing protocol are illustrated in Fig.(\ref{fig1}). Panel~(a) demonstrates the theoretical scaling relationships, contrasting the classical standard quantum limit ($\Delta \epsilon \propto 1/\sqrt{N}$) with the quantum Heisenberg limit ($\Delta \epsilon \propto 1/N$) achieved through multipartite entanglement. This quadratic improvement represents the maximum possible quantum advantage allowed by quantum mechanics, transforming the scaling from inverse square root to inverse linear with qubit number $N$.

Panel~(b) translates these scaling laws into physical sensitivity using experimentally realizable parameters from state-of-the-art superconducting quantum hardware~\cite{grabovskij2012strain, koch2007charge, blais2004cavity}. Starting from a single-qubit sensitivity of 64~ps/$\sqrt{\text{Hz}}$, the protocol achieves 6.4~ps/$\sqrt{\text{Hz}}$ with $N=10$ qubits prepared in a GHZ state (green circle), matching the performance target derived in the following section of our work. The Heisenberg scaling enables rapid progression toward the femtostrain regime, with 100 entangled qubits theoretically achieving 0.64~ps/$\sqrt{\text{Hz}}$ sensitivity.

This enhanced scaling directly results from the quadratic growth of QFI ($F_Q \propto N^2$) for entangled states, as derived in the current section. The practical implementation leverages established superconducting qubit parameters, including strain susceptibility $\chi_\epsilon = 50$~MHz/strain~\cite{grabovskij2012strain}, qubit-resonator coupling $g_0/2\pi = 50$~MHz~\cite{koch2007charge}, and interaction times $\tau = 100$~ns compatible with typical resonator decay rates~\cite{blais2004cavity}. This combination of fundamental quantum advantage and experimental feasibility positions our protocol as a viable path toward unprecedented strain sensitivity for quantum hardware diagnostics and nanoscale material characterization.

\section{Numerical Example (Picostrain)}

To provide a quantitative estimate of the achievable sensitivity, we evaluate Eq.(\ref{strain uncertainty}) using experimental parameters from state-of-the-art superconducting hardware. The strain-dependent coupling gradient $g_1 = \partial g_\epsilon / \partial \epsilon$ is the critical parameter. For a transmon qubit with capacitive coupling to a resonator, $g_1 \propto \sqrt{\omega_q \, \omega_r}$. Under strain modulation, $\omega_q (\epsilon) = \omega_q^0 + \chi_\epsilon \epsilon$, where $\chi_\epsilon$ is the strain susceptibility. This leads to a first-order approximation for the coupling gradient $g_1 \approx g_0 / (2 \, \omega_q^0) \, \chi_\epsilon$, where $g_0$ is the nominal qubit-resonator coupling at zero strain. We adopt the following values: a qubit transition frequency $\omega_q^0 / 2\pi = 5$~GHz, a strain susceptibility $\chi_\epsilon = 50$~MHz/strain (a value measured for aluminum Josephson junctions)~\cite{grabovskij2012strain}, a qubit-resonator coupling $g_0 / 2\pi = 50$~MHz, an interaction time $\tau = 100$~ns (compatible with typical resonator decay times $\kappa^{-1} > 1~\mu$s)~\cite{koch2007charge, blais2004cavity}, and a homodyne measurement at the standard quantum limit where the quadrature uncertainty $\sigma_X = 1$, corresponding to a measurement at the standard quantum limit with a quantum-efficient amplifier~\cite{blais2004cavity}.

\begin{figure*}
    \centering
    \includegraphics[width=\linewidth]{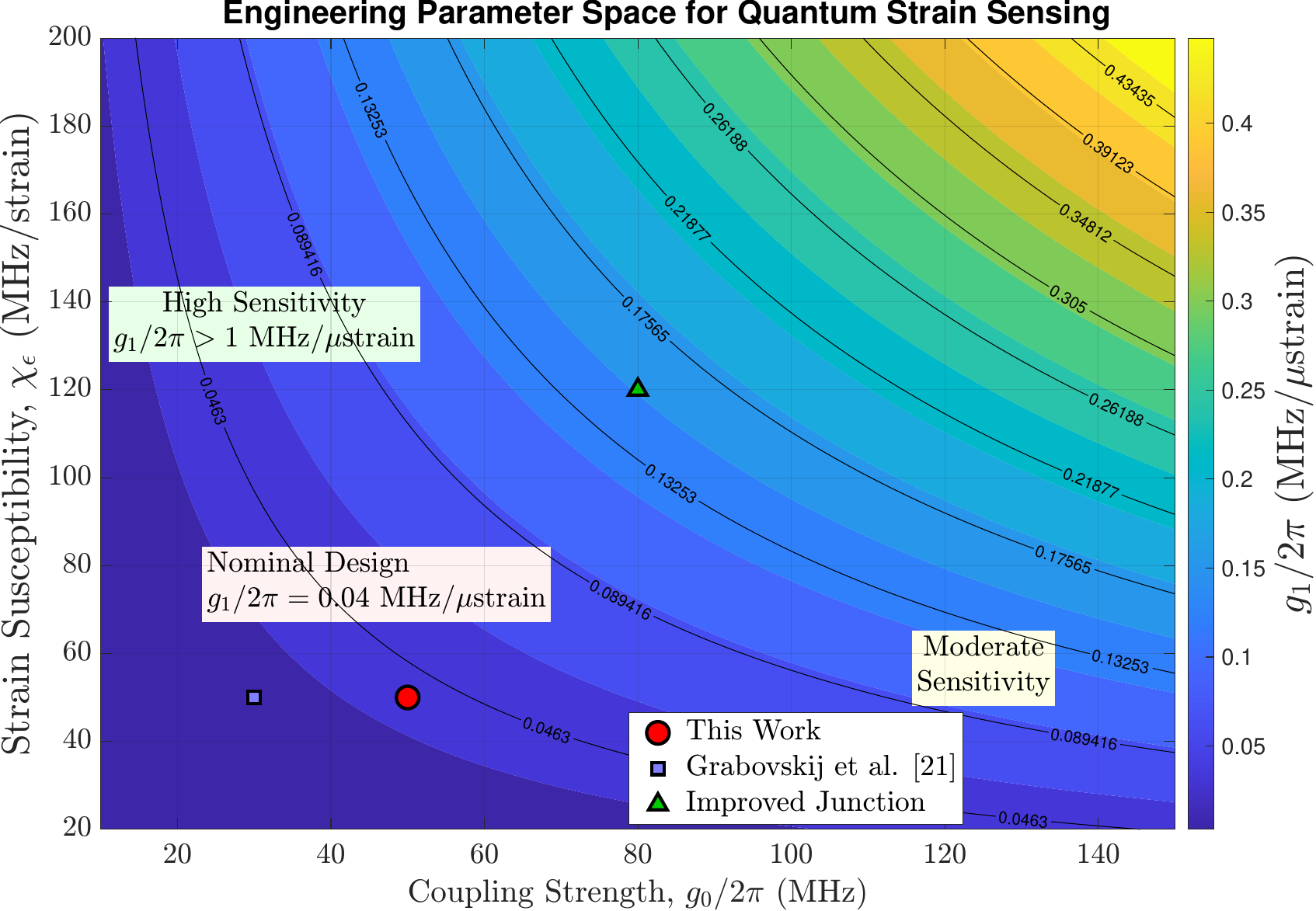}
    \caption{Engineering parameter space for the strain-dependent coupling gradient. The contour plot shows the coupling gradient \( g_{1}/2\pi \) as a function of the nominal qubit–resonator coupling strength \( g_{0}/2\pi \) and the qubit's strain susceptibility \( \chi_{\epsilon} \). The gradient \( g_{1} = (\partial g / \partial \epsilon) \) is a critical parameter determining the strain sensitivity, following the relation \( g_{1} \approx (g_{0}\chi_{\epsilon}) / (2\omega_{q0}) \). The nominal design point for this work (red circle) is compared with a state-of-the-art experimental result from Grabovskij et al.\ [20] (blue square) and a target for improved junctions (green triangle). Contours represent constant values of \( g_{1}/2\pi \) in MHz/$\mu$strain. The analysis indicates that combined optimization of \( g_{0} \) and \( \chi_{\epsilon} \) can push \( g_{1}/2\pi \) above 1~MHz/$\mu$strain, enabling enhanced picostrain and femtostrain sensitivity.
}
    \label{fig2}
\end{figure*}

Substituting these values, the coupling gradient is 
$g_1 / 2\pi \approx \frac{50~\text{MHz}}{2 \times 5~\text{GHz}} \times 50~\text{MHz/strain} = 0.25~\text{MHz}/\mu\text{strain}$. In angular frequency units, this is 
$g_1 \approx 2\pi \times 0.25 \times 10^6~\text{rad/(s$\cdot\mu$strain)} = 1.57 \times 10^6~\text{rad/(s$\cdot\mu$strain)}$. 

The single-shot strain sensitivity is then 
$\Delta \epsilon = \sigma_X / (g_1 \tau) = 1 / \big( (1.57 \times 10^6~\text{s}^{-1}) \times (100 \times 10^{-9}~\text{s}) \big) \approx 6.4 \times 10^{-3}~\text{strain}$. This corresponds to $\sim 6400~\mu\text{strain}$ per shot. By averaging over a conservative measurement bandwidth of 100~kHz ($\nu = 10^5$ independent shots per second), the sensitivity is enhanced by a factor of $1/\sqrt{\nu}$. Then, 
$\Delta \epsilon \approx (6.4 \times 10^{-3}) / \sqrt{10^5} \approx 6.4 \times 10^{-8}~\text{strain}/\sqrt{\text{Hz}} = 64~\text{picostrain}/\sqrt{\text{Hz}}$. 

While this single-qubit sensitivity is already competitive, the protocol’s full power is unlocked with entanglement. For $N$ qubits prepared in a GHZ state, the sensitivity scales as $1/N$. With a modest ensemble of $N=10$ entangled qubits, the strain resolution reaches 
$\Delta \epsilon_N \approx (64~\text{picostrain}/\sqrt{\text{Hz}})/10 \sim 6~\text{picostrain}/\sqrt{\text{Hz}}$. Further improvements using squeezed microwave states ($\sigma_X < 1$) or longer interaction times could push the sensitivity into the femtostrain regime, opening new avenues in nanoscale material science and fundamental physics.

This calculated sensitivity is directly tied to the coupling gradient \( g_{1} \). To contextualize our parameter choice and illustrate pathways for further enhancement, Fig.(\ref{fig2}) maps the engineering parameter space for \( g_{1} \). The contour plot shows \( g_{1}/2\pi \) as a function of the nominal qubit–resonator coupling strength \( g_{0}/2\pi \) and the qubit's strain susceptibility \( \chi_{\epsilon} \). Our nominal design point (red circle) is compared with a state-of-the-art experimental result (blue square) and a target for improved junctions (green triangle), illustrating a clear pathway toward enhanced picostrain and femtostrain sensitivity through combined optimization of \( g_{0} \) and \( \chi_{\epsilon} \).

\section{Applications}
The quantum strain sensing protocol developed here is not only of fundamental interest but also provides direct utility in several domains where conventional techniques fail to achieve the necessary sensitivity or integration capability. Below we highlight three representative applications. 

\subsection{In-Situ Quantum Hardware Diagnostics}

Superconducting quantum processors are highly susceptible to local strain fields generated by substrate imperfections \cite{simmonds2004decoherence, martinis2005decoherence}, packaging-induced stress, and cryogenic thermal gradients \cite{wang2023characterizing}. These strain inhomogeneities shift qubit transition frequencies \cite{grabovskij2012strain}, degrade coherence \cite{simmonds2004decoherence, martinis2005decoherence}, and cause spatial nonuniformity in multi-qubit devices. Conventional diagnostic methods lack the cryogenic compatibility and sensitivity required to map such effects at quantum scale.

Our protocol offers a natural solution, since the qubit themselves act as both the computational element and strain sensor. This dual role enables non-invasive, real-time mapping of strain fields within operating processors. Recently, coherence-stabilized qubit sensing protocols have demonstrated a 165 \% improvement in sensing efficacy compared to Ramsey interferometry without requiring additional control resources \cite{hecht2025beating}. Embedding such schemes into calibration routines can improve adaptive tuning, extend coherence lifetimes, and identify fabrication defects. Thus, the proposed approach provides an integrated pathway toward reliable-scale fault-tolerant superconducting quantum architecture.

\subsection{Nanoscale Material Characterization}

High-resolution strain mapping is central to understanding the electronic and mechanical properties of materials such as 2D superconductors \cite{ruf2021strain}, quantum heterostructures \cite{han2018strain}, and defect-engineered crystals \cite{wang2018defect}. Traditional techniques, including X-ray diffraction and electron microscopy, often fail to resolve strain gradients in operando or at the picostrain level \cite{martens2023defects}.

Quantum-enhanced approaches have already achieved notable breakthroughs. For example, nitrogen-vacancy (NV) centers in diamond have enabled ensemble-based spin interferometry with two orders of magnitude improvement in strain sensitivity, reaching microscale resolution \cite{poulsen2022optimal}. Similarly, NV-based single-qubit sensing has revealed critical spin fluctuations in magnetic thin films, opening a window into quantum phase transitions \cite{aslam2023quantum}. Our superconducting-qubit-based platform extends this paradigm to cryogenic regimes, where Josephson junction’s strain susceptibility enables characterization of superconductors and nanostructure with unprecedented precision.

\subsection{High-Frequency Gravitational Wave Detection}

Gravitational waves (GWs) in the MHz-GHz band remain experimentally unexplored, yet they are predicted to arise from early-universe phenomena and exotic particle interactions. Current interferometric detectors lack sensitivity in this regime \cite{aggarwal2021challenges}.

Superconducting quantum architectures offer a promising solution. Proposals include levitated superconducting spheres read out via flux-tunable microwave resonators, capable of achieving broadband strain sensitivity below $10^{-20}  /\sqrt (MHz) $ in the 1 kHz-1 MHz band \cite{arvanitaki2013detecting, winstone2022high}. More recently, hybrid approaches have coupled superconducting transmon qubits to high-overtone bulk acoustic resonators (hBARS), establishing a platform that could enable detection of phonon excitation induced by GHz gravitational waves or ultralight dark matter \cite{kervinen2018interfacing}. Other proposals suggest that modern quantum acoustic control could allow single-graviton detection in cooled resonators \cite{tobar2024detecting}. Our qubit-resonator strain sensing platform, with its intrinsic Heisenberg scaling, provides a natural foundation for such high-frequent GW detectors, enabling chip-scale quantum interferometry far beyond classical strain limits.

\section{Conclusions}

We present a strain sensing method that translates a qubit’s-induced frequency shift into a resonator displacement measurable via homodyne detection, achieving Heisenberg-limited picostrain sensitivity. Beyond its standalone metrological performance, a key implication of our protocol is the enabling of the quantum processor as its own diagnostic tool. The same qubits used for computation can simultaneously act as a network of ultra-sensitive strain sensors. This dual functionality allows for in-situ, real-time mapping of detrimental strain fields generated by cryogenic thermal contraction, packaging stress, or material defects. Such a capability, previously unavailable with conventional techniques, provides a direct pathway to diagnose coherence-limiting imperfections, guide improved fabrication, and implement active error mitigation, thereby accelerating the development of robust, large-scale quantum processors. This approach combines known superconducting hardware with a new sensing configuration, offering a scalable, bandwidth-rich tool for quantum technologies.

\bibliography{Ref}

@inproceedings{zhang2016high,
  title={High resolution strain sensor for earthquake precursor observation and earthquake monitoring},
  author={Zhang, Wentao and Huang, Wenzhu and Li, Li and Liu, Wenyi and Li, Fang},
  booktitle={Sixth European Workshop on Optical Fibre Sensors},
  volume={9916},
  pages={58--61},
  year={2016},
  organization={SPIE}
}

@article{gao2022review,
  title={Review of flexible piezoresistive strain sensors in civil structural health monitoring},
  author={Gao, Ke and Zhang, Zhiyue and Weng, Shun and Zhu, Hongping and Yu, Hong and Peng, Tingjun},
  journal={Applied Sciences},
  volume={12},
  number={19},
  pages={9750},
  year={2022},
  publisher={MDPI}
}

@article{ling2024novel,
  title={A novel semiconductor piezoresistive thin-film strain gauge with high sensitivity},
  author={Ling, Sijia and Chen, Xiaopeng and Miao, Lulu and Yin, Jiawen and Jian, Jiawen and Jin, Qinghui},
  journal={IEEE Sensors Journal},
  volume={24},
  number={9},
  pages={13914--13924},
  year={2024},
  publisher={IEEE}
}

@article{lyu2022strain,
  title={Strain quantum sensing with spin defects in hexagonal boron nitride},
  author={Lyu, Xiaodan and Tan, Qinghai and Wu, Lishu and Zhang, Chusheng and Zhang, Zhaowei and Mu, Zhao and Z{\'u}{\~n}iga-P{\'e}rez, Jes{\'u}s and Cai, Hongbing and Gao, Weibo},
  journal={Nano Letters},
  volume={22},
  number={16},
  pages={6553--6559},
  year={2022},
  publisher={ACS Publications}
}

@article{korkusinski2014quantum,
  title={Quantum strain sensor with a topological insulator HgTe quantum dot},
  author={Korkusinski, Marek and Hawrylak, Pawel},
  journal={Scientific reports},
  volume={4},
  number={1},
  pages={4903},
  year={2014},
  publisher={Nature Publishing Group UK London}
}

@article{yang2024strained,
  title={Strained diamond for quantum sensing applications},
  author={Yang, Limin and Wang, Heyi and Yang, Sen and Lu, Yang},
  journal={Materials for Quantum Technology},
  volume={4},
  number={2},
  pages={023001},
  year={2024},
  publisher={IOP Publishing}
}

@article{sela2020quantum,
  title={Quantum Hall response to time-dependent strain gradients in graphene},
  author={Sela, Eran and Bloch, Yakov and von Oppen, Felix and Shalom, Moshe Ben},
  journal={Physical review letters},
  volume={124},
  number={2},
  pages={026602},
  year={2020},
  publisher={APS}
}

@article{wang2023characterizing,
  title={Characterizing temperature and strain variations with qubit ensembles for their robust coherence protection},
  author={Wang, Guoqing and Barr, Ariel Rebekah and Tang, Hao and Chen, Mo and Li, Changhao and Xu, Haowei and Stasiuk, Andrew and Li, Ju and Cappellaro, Paola},
  journal={Physical Review Letters},
  volume={131},
  number={4},
  pages={043602},
  year={2023},
  publisher={APS}
}

@article{kim2023strain,
  title={Strain engineering of low-dimensional materials for emerging quantum phenomena and functionalities},
  author={Kim, Jin Myung and Haque, Md Farhadul and Hsieh, Ezekiel Y and Nahid, Shahriar Muhammad and Zarin, Ishrat and Jeong, Kwang-Yong and So, Jae-Pil and Park, Hong-Gyu and Nam, SungWoo},
  journal={Advanced Materials},
  volume={35},
  number={27},
  pages={2107362},
  year={2023},
  publisher={Wiley Online Library}
}

@article{gagliardi2010probing,
  title={Probing the ultimate limit of fiber-optic strain sensing},
  author={Gagliardi, Gianluca and Salza, M and Avino, S and Ferraro, P and De Natale, P},
  journal={Science},
  volume={330},
  number={6007},
  pages={1081--1084},
  year={2010},
  publisher={American Association for the Advancement of Science}
}

@article{giovannetti2011advances,
  title={Advances in quantum metrology},
  author={Giovannetti, Vittorio and Lloyd, Seth and Maccone, Lorenzo},
  journal={Nature photonics},
  volume={5},
  number={4},
  pages={222--229},
  year={2011},
  publisher={Nature Publishing Group UK London}
}

@article{ye2024essay,
  title={Essay: Quantum sensing with atomic, molecular, and optical platforms for fundamental physics},
  author={Ye, Jun and Zoller, Peter},
  journal={Physical Review Letters},
  volume={132},
  number={19},
  pages={190001},
  year={2024},
  publisher={APS}
}

@article{udvarhelyi2018spin,
  title={Spin-strain interaction in nitrogen-vacancy centers in diamond},
  author={Udvarhelyi, P{\'e}ter and Shkolnikov, Vladyslav O and Gali, Adam and Burkard, Guido and P{\'a}lyi, Andr{\'a}s},
  journal={Physical Review B},
  volume={98},
  number={7},
  pages={075201},
  year={2018},
  publisher={APS}
}

@article{paudel2024sensing,
  title={Sensing at the Nanoscale Using Nitrogen-Vacancy Centers in Diamond: A Model for a Quantum Pressure Sensor},
  author={Paudel, Hari P and Lander, Gary R and Crawford, Scott E and Duan, Yuhua},
  journal={Nanomaterials},
  volume={14},
  number={8},
  pages={675},
  year={2024},
  publisher={MDPI}
}

@article{roberts2025quantum,
  title={Quantum Sensing with Spin Defects Beyond Diamond},
  author={Roberts, Henry and Abudayyeh, Hamza and Li, Xiaoqin and Li, Xiuling},
  journal={ACS nano},
  year={2025},
  publisher={ACS Publications}
}

@article{maletinsky2012robust,
  title={A robust scanning diamond sensor for nanoscale imaging with single nitrogen-vacancy centres},
  author={Maletinsky, Patrick and Hong, Sungkun and Grinolds, Michael Sean and Hausmann, Birgit and Lukin, Mikhail D and Walsworth, Ronald L and Loncar, Marko and Yacoby, Amir},
  journal={Nature nanotechnology},
  volume={7},
  number={5},
  pages={320--324},
  year={2012},
  publisher={Nature Publishing Group UK London}
}

@article{kenny2025quantum,
  title={Quantum sensing enhancement through a nuclear spin register in nitrogen-vacancy centers in diamond},
  author={Kenny, Jonathan and Zhou, Feifei and He, Ruihua and Jelezko, Fedor and Koh, Teck Seng and Gao, Weibo},
  journal={Applied Physics Reviews},
  volume={12},
  number={2},
  year={2025},
  publisher={AIP Publishing}
}

@article{simmonds2004decoherence,
  title={Decoherence in Josephson phase qubits from junction resonators},
  author={Simmonds, Raymond W and Lang, KM and Hite, Dustin A and Nam, S and Pappas, David P and Martinis, John M},
  journal={Physical Review Letters},
  volume={93},
  number={7},
  pages={077003},
  year={2004},
  publisher={APS}
}

@article{martinis2005decoherence,
  title={Decoherence in Josephson qubits from dielectric loss},
  author={Martinis, John M and Cooper, Ken B and McDermott, Robert and Steffen, Matthias and Ansmann, Markus and Osborn, KD and Cicak, Katarina and Oh, Seongshik and Pappas, David P and Simmonds, Raymond W and others},
  journal={Physical review letters},
  volume={95},
  number={21},
  pages={210503},
  year={2005},
  publisher={APS}
}

@article{ccelik2025quantum,
  title={Quantum interferometric protocol using spin-dependent displacements},
  author={{\c{C}}elik, Necati and Akbulut {\"O}zen, Song{\"u}l and Engin, Burhan},
  journal={Physical Review A},
  volume={112},
  number={4},
  pages={042429},
  year={2025},
  publisher={APS}
}

@article{grabovskij2012strain,
  title={Strain tuning of individual atomic tunneling systems detected by a superconducting qubit},
  author={Grabovskij, Grigorij J and Peichl, Torben and Lisenfeld, J{\"u}rgen and Weiss, Georg and Ustinov, Alexey V},
  journal={Science},
  volume={338},
  number={6104},
  pages={232--234},
  year={2012},
  publisher={American Association for the Advancement of Science}
}

@article{cao2024photonic,
  title={Photonic source of heralded Greenberger-Horne-Zeilinger states},
  author={Cao, H and Hansen, LM and Giorgino, F and Carosini, L and Zah{\'a}lka, P and Zilk, F and Loredo, JC and Walther, P},
  journal={Physical Review Letters},
  volume={132},
  number={13},
  pages={130604},
  year={2024},
  publisher={APS}
}

@article{koch2007charge,
  title={Charge-insensitive qubit design derived from the Cooper pair box},
  author={Koch, Jens and Yu, Terri M and Gambetta, Jay and Houck, Andrew A and Schuster, David I and Majer, Johannes and Blais, Alexandre and Devoret, Michel H and Girvin, Steven M and Schoelkopf, Robert J},
  journal={Physical Review A—Atomic, Molecular, and Optical Physics},
  volume={76},
  number={4},
  pages={042319},
  year={2007},
  publisher={APS}
}

@article{zallo2014strain,
  title={Strain-induced active tuning of the coherent tunneling in quantum dot molecules},
  author={Zallo, Eugenio and Trotta, R and K{\v{r}}{\'a}pek, V and Huo, YH and Atkinson, Paola and Ding, F and {\v{S}}ikola, T and Rastelli, A and Schmidt, OG},
  journal={Physical Review B},
  volume={89},
  number={24},
  pages={241303},
  year={2014},
  publisher={APS}
}

@article{faoro2005models,
  title={Models of environment and T 1 relaxation in Josephson charge qubits},
  author={Faoro, Lara and Bergli, Joakim and Altshuler, Boris L and Galperin, Yuri M},
  journal={Physical review letters},
  volume={95},
  number={4},
  pages={046805},
  year={2005},
  publisher={APS}
}

@article{blais2004cavity,
  title={Cavity quantum electrodynamics for superconducting electrical circuits: An architecture for quantum computation},
  author={Blais, Alexandre and Huang, Ren-Shou and Wallraff, Andreas and Girvin, Steven M and Schoelkopf, R Jun},
  journal={Physical Review A—Atomic, Molecular, and Optical Physics},
  volume={69},
  number={6},
  pages={062320},
  year={2004},
  publisher={APS}
}

@article{helstrom1969quantum,
  title={Quantum detection and estimation theory},
  author={Helstrom, Carl W},
  journal={Journal of Statistical Physics},
  volume={1},
  number={2},
  pages={231--252},
  year={1969},
  publisher={Springer}
}

@article{hecht2025beating,
  title={Beating the Ramsey limit on sensing with deterministic qubit control},
  author={Hecht, MO and Saurav, Kumar and Vlachos, Evangelos and Lidar, Daniel A and Levenson-Falk, Eli M},
  journal={Nature Communications},
  volume={16},
  number={1},
  pages={3754},
  year={2025},
  publisher={Nature Publishing Group UK London}
}

@article{ruf2021strain,
  title={Strain-stabilized superconductivity},
  author={Ruf, Jacob P and Paik, Hanjong and Schreiber, Nathaniel J and Nair, Hari P and Miao, Ludi and Kawasaki, Jason K and Nelson, Jocienne N and Faeth, Brendan D and Lee, Yonghun and Goodge, Berit H and others},
  journal={Nature Communications},
  volume={12},
  number={1},
  pages={59},
  year={2021},
  publisher={Nature Publishing Group UK London}
}

@article{han2018strain,
  title={Strain mapping of two-dimensional heterostructures with subpicometer precision},
  author={Han, Yimo and Nguyen, Kayla and Cao, Michael and Cueva, Paul and Xie, Saien and Tate, Mark W and Purohit, Prafull and Gruner, Sol M and Park, Jiwoong and Muller, David A},
  journal={Nano letters},
  volume={18},
  number={6},
  pages={3746--3751},
  year={2018},
  publisher={ACS Publications}
}

@article{wang2018defect,
  title={Defect-engineered epitaxial VO2$\pm$$\delta$ in strain engineering of heterogeneous soft crystals},
  author={Wang, Yiping and Sun, Xin and Chen, Zhizhong and Cai, Zhonghou and Zhou, Hua and Lu, Toh-Ming and Shi, Jian},
  journal={Science advances},
  volume={4},
  number={5},
  pages={eaar3679},
  year={2018},
  publisher={American Association for the Advancement of Science}
}

@article{martens2023defects,
  title={Defects and nanostrain gradients control phase transition mechanisms in single crystal high-voltage lithium spinel},
  author={Martens, Isaac and Vostrov, Nikita and Mirolo, Marta and Leake, Steven J and Zatterin, Edoardo and Zhu, Xiaobo and Wang, Lianzhou and Drnec, Jakub and Richard, Marie-Ingrid and Schulli, Tobias U},
  journal={Nature Communications},
  volume={14},
  number={1},
  pages={6975},
  year={2023},
  publisher={Nature Publishing Group UK London}
}

@article{poulsen2022optimal,
  title={Optimal control of a nitrogen-vacancy spin ensemble in diamond for sensing in the pulsed domain},
  author={Poulsen, Andreas FL and Clement, Joshua D and Webb, James L and Jensen, Rasmus H and Troise, Luca and Berg-S{\o}rensen, Kirstine and Huck, Alexander and Andersen, Ulrik Lund},
  journal={Physical Review B},
  volume={106},
  number={1},
  pages={014202},
  year={2022},
  publisher={APS}
}

@article{aslam2023quantum,
  title={Quantum sensors for biomedical applications},
  author={Aslam, Nabeel and Zhou, Hengyun and Urbach, Elana K and Turner, Matthew J and Walsworth, Ronald L and Lukin, Mikhail D and Park, Hongkun},
  journal={Nature Reviews Physics},
  volume={5},
  number={3},
  pages={157--169},
  year={2023},
  publisher={Nature Publishing Group UK London}
}

@article{aggarwal2021challenges,
  title={Challenges and opportunities of gravitational-wave searches at MHz to GHz frequencies},
  author={Aggarwal, Nancy and Aguiar, Odylio D and Bauswein, Andreas and Cella, Giancarlo and Clesse, Sebastian and Cruise, Adrian Michael and Domcke, Valerie and Figueroa, Daniel G and Geraci, Andrew and Goryachev, Maxim and others},
  journal={Living reviews in relativity},
  volume={24},
  number={1},
  pages={4},
  year={2021},
  publisher={Springer}
}

@article{arvanitaki2013detecting,
  title={Detecting high-frequency gravitational waves with optically levitated sensors},
  author={Arvanitaki, Asimina and Geraci, Andrew A},
  journal={Physical review letters},
  volume={110},
  number={7},
  pages={071105},
  year={2013},
  publisher={APS}
}

@inproceedings{winstone2022high,
  title={High frequency gravitational wave detection with optically levitated nanoparticles: an update on LSD (levitated sensor detector)},
  author={Winstone, George Paul and Grass, Daniel H and Wang, Aron and Klomp, Shelby and Laeuger, Andrew and Galla, Chethn and Montoya, Cris and Aggarwal, Nancy and Sprague, Jacob and Poverman, Andrew and others},
  booktitle={Optical and Quantum Sensing and Precision Metrology II},
  volume={12016},
  pages={57--66},
  year={2022},
  organization={SPIE}
}

@article{kervinen2018interfacing,
  title={Interfacing planar superconducting qubits with high overtone bulk acoustic phonons},
  author={Kervinen, Mikael and Rissanen, Ilkka and Sillanp{\"a}{\"a}, Mika},
  journal={Physical Review B},
  volume={97},
  number={20},
  pages={205443},
  year={2018},
  publisher={APS}
}

@article{tobar2024detecting,
  title={Detecting single gravitons with quantum sensing},
  author={Tobar, Germain and Manikandan, Sreenath K and Beitel, Thomas and Pikovski, Igor},
  journal={Nature Communications},
  volume={15},
  number={1},
  pages={7229},
  year={2024},
  publisher={Nature Publishing Group UK London}
}

\end{document}